\documentclass[english]{IEEEtran}
\usepackage[T1]{fontenc}
\usepackage[utf8]{inputenc}
\usepackage{array}
\usepackage{amssymb}
\usepackage{graphicx}
\usepackage{url}
\usepackage{amsmath}
\usepackage{balance}

\makeatletter

\providecommand{\tabularnewline}{\\}

\usepackage[caption=true,font=footnotesize]{subfig}
\usepackage{algorithm}
\usepackage{algorithmic}[1]
\makeatother

\usepackage{babel}
\begin{document}

\title{A Localization Approach for Crowdsourced Air Traffic Communication Networks}

\author{Martin Strohmeier$^{*}$, Vincent Lenders$^{+}$, Ivan Martinovic$^{*}$\\
$^{*}$University of Oxford, United Kingdom $^{+}$armasuisse, Switzerland\\
\{martin.strohmeier\}, \{ivan.martinovic\}@cs.ox.ac.uk, vincent.lenders@armasuisse.ch}
\maketitle
\begin{abstract}
In this work, we argue that current state-of-the-art methods of aircraft localization such as multilateration are insufficient, in particular for modern crowdsourced air traffic networks with random, unplanned deployment geometry. We propose an alternative, a grid-based localization approach using the k-Nearest Neighbor algorithm, to deal with the identified shortcomings. Our proposal does not require any changes to the existing air traffic protocols and transmitters, and is easily implemented using only low-cost, commercial-off-the-shelf hardware.

Using an algebraic multilateration algorithm for comparison, we evaluate our approach using real-world flight data collected with our collaborative sensor network OpenSky. We quantify its effectiveness in terms of aircraft location accuracy, surveillance coverage, and the verification of false position data.

Our results show that the grid-based approach can increase the effective air traffic surveillance coverage compared to multilateration by a factor of up to 2.5. As it does not suffer from dilution of precision, it is much more robust in noisy environments and performs better in pre-existing, unplanned receiver deployments. We further find that the mean aircraft location accuracy can be increased by up to 41\% in comparison with multilateration while also being able to pinpoint the origin of potential spoofing attacks conducted from the ground.

\end{abstract}

\section{Introduction}

Air traffic control (ATC) is the backbone of what is arguably the key means of personal transport in the modern world. As traffic continues to grow dramatically, ATC has to manage ever more aircraft. Large European airports, such as London Heathrow or Frankfurt/Main, experience spikes of more than 1,500 daily take-offs and landings, and industry forecasts predict that world-wide flight movements will double between 2015 and 2034 \cite{Boeing}. Additionally, as Unmanned Aerial Vehicles (UAV) enter the civil airspace, they must learn to co-exist with manned aircraft and existing air traffic control systems. While forecasts project a steady 5\% annual increase of global manned flight traffic over the next 20 years, UAV are projected to outgrow traditional air traffic by several orders of magnitude: In 2035, 250,000 UAV are expected to be operating in the US alone, compared to a mere 45,000 passenger aircraft around the globe \cite{UAV}. However, as this paradigm shift progresses, many technological issues have yet to be solved to ensure the safe control of both manned and unmanned aircraft. 

The key issue in controlling the airspace is to know where an aircraft is at any given time. For the pilot, this is achieved through navigational aids, such as Distance Measuring Equipment (DME) or, more recently, satellite navigation. For ground controllers, the traditional options to obtain an aircraft's position comprise voice communication via VHF/HF and primary and secondary surveillance radar (PSR and SSR, respectively). Recently, technological developments and stricter separation needs have given rise to other methods of aircraft localization, most notably the the Automatic Dependent Broadcast--Protocol (ADS-B) and multilateration (MLAT).

While ADS-B relies purely on the aircraft correctly broadcasting their own position to other aircraft and ground stations, MLAT can provide an independent means of localization by exploiting the time differences of arrival (TDoA) of signals received at several different ground stations. Both methods achieve greatly increased localization accuracy compared to previous radar surveillance and are key technologies in next generation ATC concepts such as SESAR or NextGen \cite{Strohmeier2016}.

At the same time, crowdsourced air traffic communication networks have gained importance over the past decade. Large private companies such as FlightRadar24 and PlaneFinder,\footnote{http://flightradar24.com, http://planefinder.net} research networks such as OpenSky,\footnote{http://opensky-network.org} enthusiast websites,\footnote{http://adsbexchange.com} and increasingly flight authorities themselves \cite{ReutersGermanWings} rely less and less on planned deployments of ATC receivers. Instead they use distributed networks that are randomly deployed from the local to the global level. Contrary to traditional, carefully planned receiver networks, this crowdsourced use of cheap sensors provides a number of new challenges to the existing localization algorithms.

In this work, we investigate alternatives to multilateration for independent localization in air traffic communication networks. We present a new method based on expected TDoAs, which is robust in noisy environments and does not depend on the system's receiver geometry. We show that it is effective even in unplanned deployments of cheap, crowd-sourced, off-the-shelf ATC receiver networks. Our approach does not require changes to existing technology standards or to the aircraft's legacy hardware equipment, which is particularly important given aviation's long adoption and certification cycles \cite{strohmeier2016cycon}.

We make the following contributions in this paper:\\
\vspace{-5pt}

\begin{itemize}
\item We develop a new grid-based method to localize aircraft based on their wireless communication. We utilize a new combination of the k-Nearest Neighbor algorithm and the expected time differences of arrival of ATC signals to estimate their origin.\\

\item We analyse the disadvantages of multilateration, the current state of the art solution for aircraft localization. Concretely, we examine the robustness in noisy environments, the coverage in a typical air traffic surveillance area, and the efficiency of the signal usage.\\

\item We evaluate our approach on real-world data and show that it performs better in wide area settings than multilateration. Compared to the latter, our approach is cheaper and more scalable, and improves surveillance range, detection speed and location accuracy in real-world environments. In particular, our approach is much less dependent on perfect system geometry.\\

\item Finally, we study the effectiveness of verifying air traffic control data. We show that verification of legitimate and false flight data using our approach is quicker and can also provide improved localization of the origin of false data injections on or near the ground.
\end{itemize}

We have	introduced the core localization idea in the 1st ACM Workshop on Cyber-Physical System	Security (CPSS '15), using it for security and data verification purposes only \cite{strohmeier2015lightweight}. This journal version is fully rewritten, focusing on the localization part and providing extensive analysis concerning the benefits in modern crowdsourced air traffic communication sensor networks. We have provided new data and new findings to substantiate our approach, including the theoretical underpinning of expected TDoAs, a robustness analysis, a new discussion of its usefulness for verification, and extensive comparisons with multilateration in crowdsourced networks.

The remainder of this paper is organized as follows. Section \ref{sec:Background} discusses the background necessary for the aircraft localization problem, including the related work. Section \ref{sec:Multilateration} analyzes the problems of multilateration while Section \ref{sec:TDOA-Coverage} illustrates the differences between TDoA-based algorithms. Section \ref{sec:design} explains the concept of expected TDoAs, and how it can be exploited for lightweight aircraft localization. Section \ref{sec:Experimental-Design} details our experimental setup, whereas Section \ref{sec:Evaluation} evaluates the scheme against real-world flight data. Section \ref{sec:attack-detection} examines the scheme's data verification effectiveness. Finally, Section \ref{sec:Conclusion} summarizes and concludes this work. 

\section{Background}\label{sec:Background}

\subsection{Modern Air Traffic Control}

\begin{figure}
\centering
\includegraphics[scale=0.49]{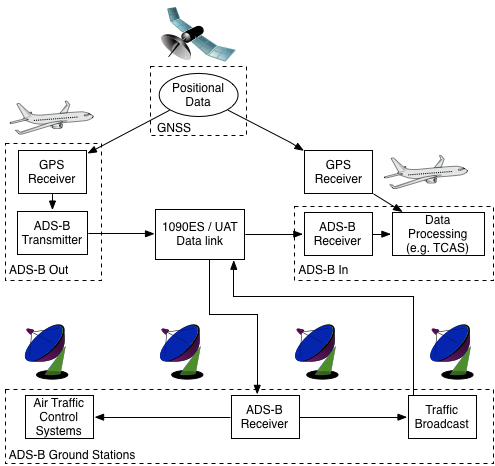}\caption[Overview of the ADS-B system architecture.]{Overview of the ADS-B system architecture. Aircraft receive positional data that is transmitted via the ADS-B Out subsystem over the 1090ES or the UAT data link. It is then received and processed by ground stations and by other aircraft via the ADS-B In subsystem. \label{fig:Overview-ADS-B}}
\end{figure}

Since World War II, ATC surveillance data has been provided by PSR and SSR in various forms. PSR is a term comprising non-cooperative radar localization systems that use a rotating antenna radiating a pulse position-modulated and highly directional electromagnetic beam on a low GHz band. The pulses are returned by all reflecting targets in the airspace; measuring the bearing and round trip time provides the targets' positions. 

In contrast, SSR is a cooperative technology comprising the so-called transponder Modes A, C, and S. Ground stations interrogate the transponders of aircraft in range using digital messages on the 1030 MHz frequency, which reply with the requested information on the 1090 MHz channel. SSR provides additional target information compared to PSR besides the position, including identity and altitude.

One of the most recently introduced ATC technologies is the ADS-B  protocol, which is the satellite-based successor of SSR. It is currently being rolled out in most airspaces and promises lower cost and more accurate surveillance~\cite{FederalAviationAdministration2016}. Contrary to primary and secondary radar, it does not rely on interrogations. Using two potential datalinks (1090 Extended Squitter, or Universal Access Transceiver), the aircraft automatically broadcasts its position and velocity (twice per second) as well as its identification (every five seconds) to all other aircraft and ground stations in the vicinity (see Fig. \ref{fig:Overview-ADS-B} for an illustration). 

As it is possible to capture all popular wireless protocols with cheap off-the-shelf wireless receivers, crowdsourced sensor networks have increased in popularity over recent years \cite{strohmeier2016cycon}. Using any wireless receiver such as popular software-defined radios, any signal sent by the aircraft, including ADS-B and SSR, can be used for TDoA-based localization such as multilateration \cite{Schaefer14}.

\subsection{Characteristics of Aircraft Localization}

To be able to develop an appropriate localization method, in particular for crowdsourced networks, it is crucial to first identify the characteristics of the ATC environment. The following characteristics distinguish the aircraft localization problem from other wireless localization problems (e.g., wireless sensor networks or vehicular ad hoc networks):
\begin{itemize}
\item \textbf{Outdoor line-of-sight environment:} Contrary to many localization problems found in academic research, the aircraft location problem is naturally outdoors. On the 1090\,MHz channel, the line of sight (LoS) is a crucial factor in receiving signals. We require an outdoor LoS propagation model for our work in terms of loss and propagation.
\item \textbf{Vast distances:} In \emph{wide area} ATC surveillance, the distances covered are naturally much larger than in more local or indoor problems. Aircraft flying at cruising altitudes (typically 35,000 feet or higher for commercial aircraft) can be observed up to the radio horizon of 400\,km or more. This is orders of magnitude larger than typical indoor localization problems.
\item \textbf{Few multipath effects:} At typical aircraft cruising altitudes, we experience comparably few diffractions leading to multipath effects that influence signal characteristics. This enables us to use simpler theoretical models than in more complex indoor and multipath-rich environments. Most importantly, the propagation timings between aircraft positions and sensors can be approximated easily by using the speed of light $c$.
\end{itemize}

\subsection{Related Work}

Indoor and outdoor localization problems have been studied extensively in the literature, often in the scope of sensor networks and radar applications. Liu et al.\,\cite{liu2007survey} give an overview of the techniques used in wireless indoor positioning including the different algorithms (k-Nearest Neighbor, lateration, least squares and Bayesian among others) and primitives such as received signal strength (RSS), TDoA, time of arrival (ToA) and angle of arrival (AoA). RSS-based methods are the most popular within any type of wireless networks as they are often readily supported out of the box and do not require additional hardware such as high precision clocks or antenna arrays.

While TDoA systems are limited in indoor environments (due to multipath effects and non-availability of time synchronization and clocks fine-grained enough to provide good results at very short distances \cite{bahl2000radar}), they offer very superior performance in long-distance outdoor environments such as those encountered in the aircraft localization problem where the non-line-of-sight error is not the dominant error source \cite{li2005method}. Errors in RSS-based outdoor localization are typically much larger, ranging up to a few hundred meters already in comparatively small outdoor settings of 2\,km \cite{ning2016outdoor}.

In terms of algorithms, the k-Nearest Neighbors (k-NN) algorithm has proven to do very well in short-distance, indoor RSS fingerprinting compared to other methods \cite{rozyyev2012combined}, although it has been studied less in long-distance scenarios (such as aircraft localization) and can become computationally more expensive with very large databases. 

\begin{figure}
\begin{centering}
\includegraphics[scale=0.5]{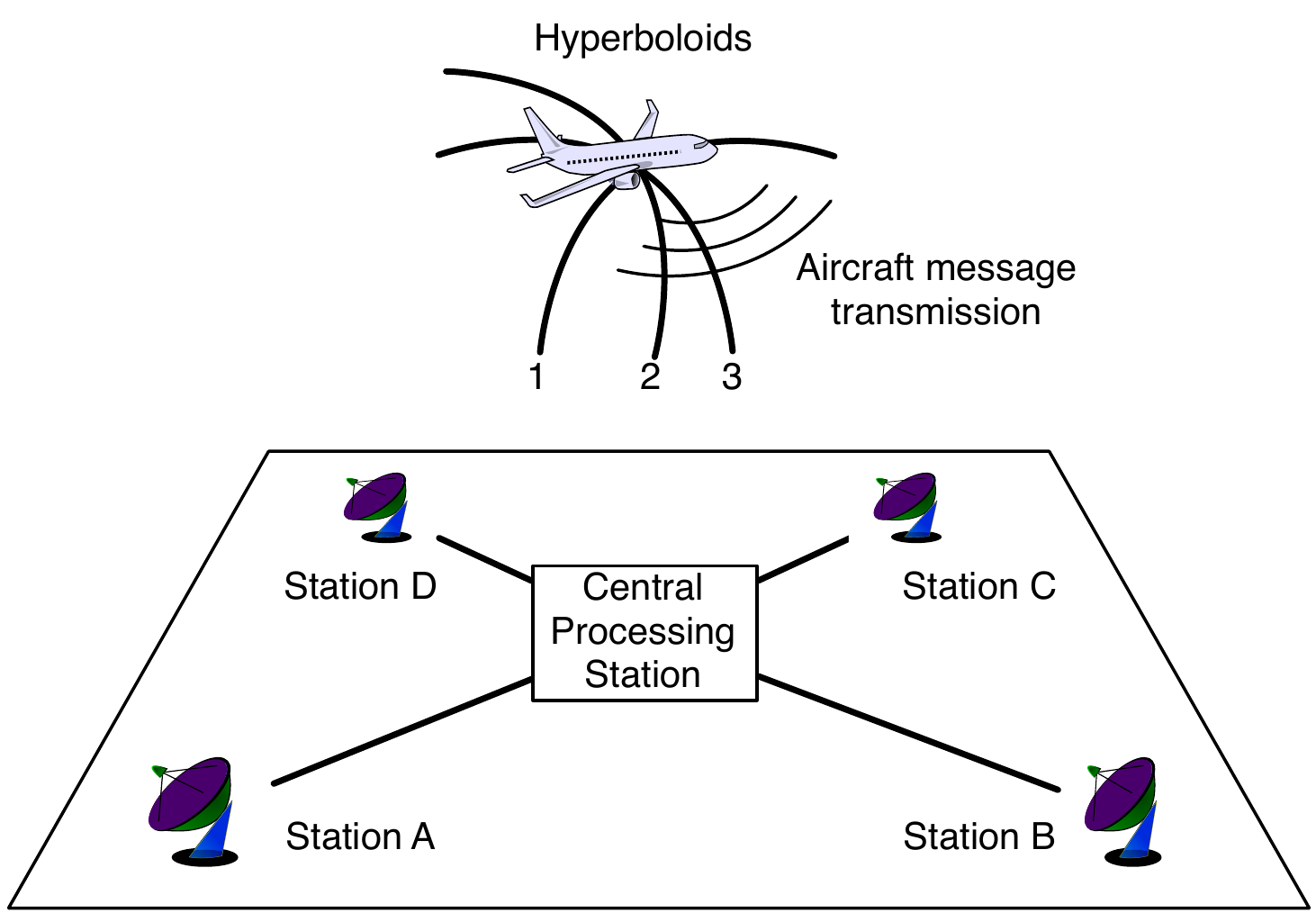}
\par\end{centering}

\caption[Basic multilateration architecture.]{Basic multilateration architecture with four receivers and three resulting hyperboloids.\label{fig:MLAT-architecture.}}
\end{figure}

Overall, the main (distributed) localization approach used within aviation is MLAT \cite{strohmeier2015lightweight}, which we discuss in detail in the next section and which provides the baseline for our evaluations. To the best of our knowledge the combination of the k-NN algorithm with TDoA as a primitive has not been studied. In the following, we argue that this combination is beneficial in particular for crowdsourced networks with random, imperfect system geometry in which the performance of MLAT suffers strongly.

\section{Multilateration}\label{sec:Multilateration}

MLAT is a proven and well-understood concept that is used in civil and military navigation and already serves as a backup for ATC around some airports. It has been the consensus solution in academia and aviation circles as a backup solution for primary and secondary radar \cite{Strohmeier2016}. 

\subsection{Concept}

To localize the 2D-origin of a signal using MLAT, three (or more) receiver stations measure the time at which they receive the same message from an aircraft. They send this data to the central processing station which can calculate the aircraft's position from the intersecting hyperboloids that result from the time difference of arrival between the receiver stations. 

Traditionally, the algorithms to solve this problem are classified into open form (or iterative) and closed form (or direct). This classification of algorithms is based on their need of information from external sources: those that require a previous estimation of the solution are considered open form and those that do not require such information are closed form. 

Mantilla-Gaviria et al.\,\cite{mantilla2015localization} discuss the different MLAT models used for airport surface surveillance and classify them into three main categories: numerical, statistical and algebraic. The resulting problem obtained by these models is then numerically solved, e.g., by using least squares or singular value decomposition-based regularization.

\subsection{Downsides}

However, despite its popularity, MLAT suffers from a number of known pitfalls.

\begin{figure}
\begin{centering}
\includegraphics[width=0.7\columnwidth]{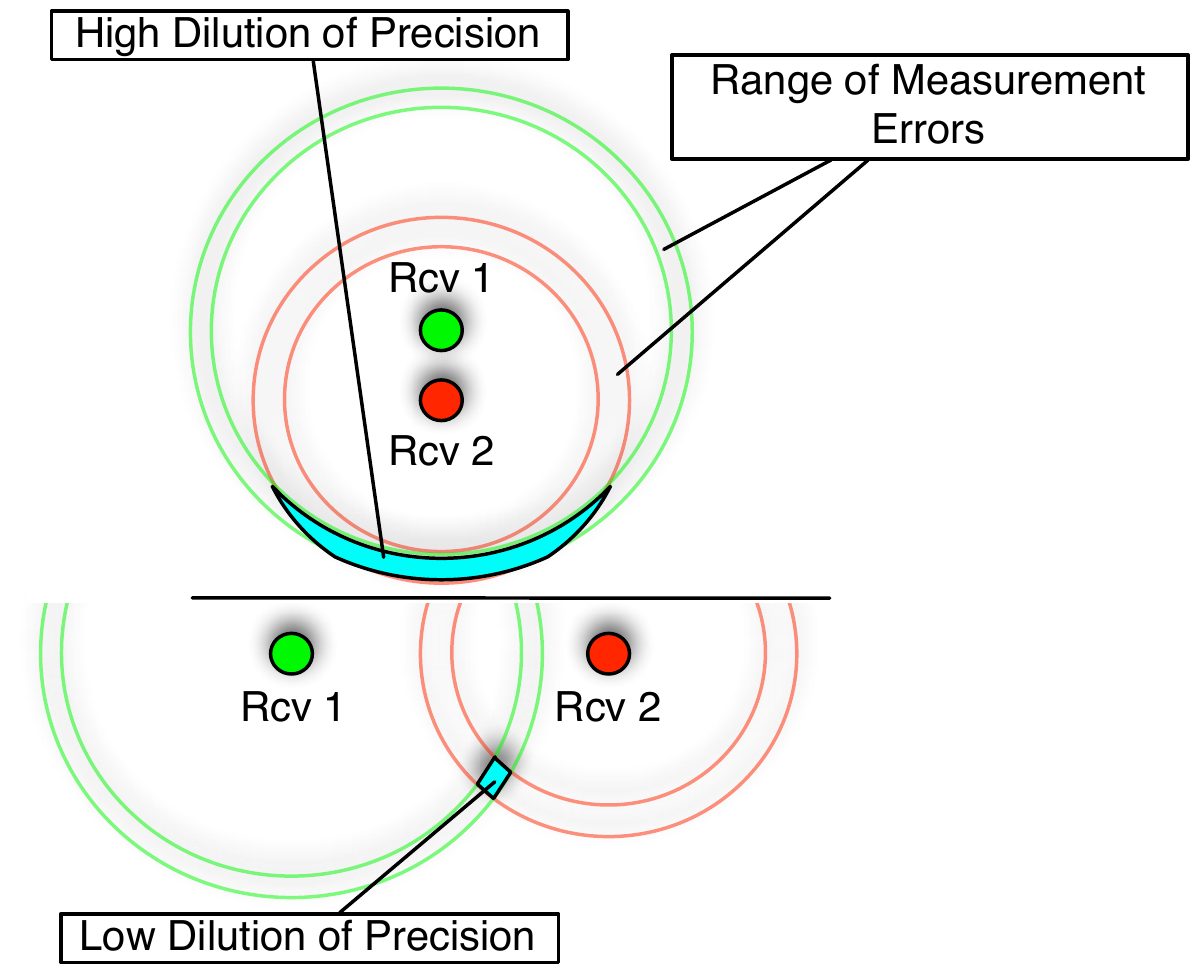}
\par\end{centering}

\protect\caption[Illustration of geometric dilution of precision.]{Geometric dilution of precision. The circles show the measurement errors of the respective receivers; the intersections demonstrate the area where the true location of the measured object can be found. An adverse placement of receivers in relation to the target (top) can severely affect the outcome of the localization compared to a favorable deployment (bottom).
\label{fig:Geometric-dilution-of}}
\end{figure}

First, MLAT is highly susceptible to noise, outliers, and even minor measurement errors outside a small core area. An important quality metric for a deployment and its MLAT accuracy with respect to the target object's relative position is the \emph{geometric dilution of precision}, or GDOP (see Fig.\,\ref{fig:Geometric-dilution-of}). It describes the effect of a deployment on the relationship between the errors of the obtained TDoA measurements and their resulting impact on the final errors in the object's calculated position, or formally:
\begin{equation}
\Delta Location\, Estimate=\Delta Measurements\,\cdot\, GDOP
\end{equation}
GDOP is widely used in positioning systems such as GPS, where good ratings for this multiplier are commonly considered to be below 6, with 10 to be fair and everything over 20 to be of poor quality \cite{mosavi2011applying}. For a full explanation of how to calculate different DOP values, please see \cite{langley1999dilution}.\\

Second, traditional MLAT systems are very expensive: Where ADS-B needs only a single receiver for accurate wide area surveillance of up to 400\,km, MLAT requires every signal to be received by at least three stations with little noise. On top of this, geographical obstacles (e.g., mountain ranges, oceans) make it relatively more difficult to install a comprehensive wide area system at the desired service level.
However, even where this is not a problem, MLAT's discussed dilution of precision characteristics necessitate a well-planned system geometry to accurately cover a given area \cite{mantilla2015localization}; an unplanned, or crowdsourced sensor network deployment cannot typically provide good enough GDOP values to achieve the same service quality.
\\

Considering these drawbacks and the fact that modern crowdsourced networks with cheap off-the shelf sensors are becoming more popular, there is a need for other TDoA-based approaches that can improve on these problems.

\begin{figure}
\begin{centering}
\includegraphics[width=1\columnwidth]{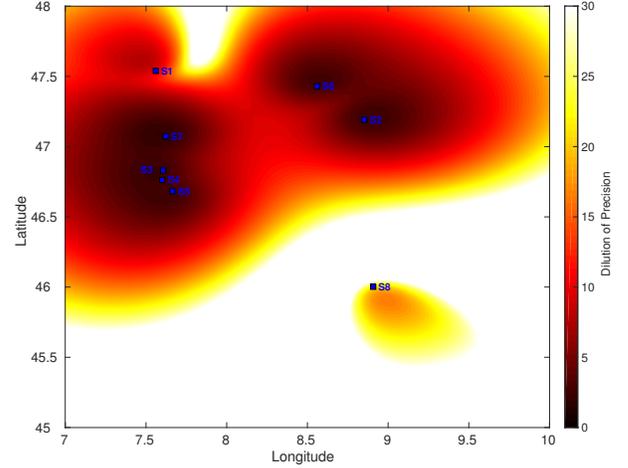}
\par\end{centering}

\protect\caption[Map of the practical reception ranges of an 8 sensor ADS-B system.]{The map shows the geometric dilution of precision of our 8 sensor measurement system for wide area multilateration. All sensors are on the ground, the assumed measurement altitude is 38,000 feet. \label{fig:dop-map}}
\end{figure}

\section{Analysing Scalability and Coverage of Crowdsourced Localization}\label{sec:TDOA-Coverage}

One of our main goals is to tackle MLAT's scalability and coverage problems, especially in deployments with less than perfect system geometry. An ATC data communications network consists of a given number of sensors that are deployed outside, in a line of sight with the airspace they are expected to cover. Naturally, overlapping reception ranges between receivers are required to obtain TDoAs. If more sensors are to receive the same message, they need to be located closer together. While this increases the overlap, it also decreases the overall ADS-B coverage of the receivers. Worse even, only a small part of the MLAT coverage is usable, since GDOP causes its accuracy to deteriorate quickly. Methods not suffering from GDOP and working with fewer sensors could vastly improve efficiency compared to MLAT.

Figure \ref{fig:dop-map} illustrates the dilution of precision experienced by multilateration in our crowdsourced deployment using 8 ADS-B sensors. Covering an area of 3 degrees longitude $\times$ 3 degrees latitude at an altitude of 38,000 feet, it is indicative of a wide area multilateration system used for en-route airspace surveillance. As can be observed, the darkest areas, which illustrate the lowest GDOP values, only make up a fraction of the surveillance area, while the system's geometry causes most parts to have unusably high dilution of precision.  

\begin{table}
\begin{centering}
\begin{tabular}{|>{\centering}m{2.1cm}|>{\centering}m{1.35cm}|c||c|}
\hline 
 & Absolute  & Relative & Area covered\tabularnewline
\hline 
All messages & 63,410,017 & 100\% & 100\%\tabularnewline
\hline 
\# seen by >=2 sensors & 26,305,043 & 41.48\% & 46.51\%\tabularnewline
\hline 
\# seen by >=3 sensors & 8,912,305 & 14.06\% & 17.25\%\tabularnewline
\hline 
\# seen by >=4 sensors & 2,560,727 & 4.04\% & 5.28\%\tabularnewline
\hline 
\# seen by >=5 sensors & 394,362 & 0.62\% & 0.84\%\tabularnewline
\hline 
\# seen by >=6 sensors & 18,777 & 0.0003\% & 0.0004\%\tabularnewline
\hline 
\# seen by >=7 sensors & 97 & $1.5*10^{-6}$\% & $2.1*10^{-6}$\%\tabularnewline
\hline 
\hline 
\# MLAT \& GDOP < 10 & 3,319,618 & 5.24\% & 6.67\%\tabularnewline
\hline 
\end{tabular}
\par\end{centering}

\protect\caption[Statistics on OpenSky dataset used for aircraft localization.]{Statistics on the utilized OpenSky dataset. The table shows the absolute and relative number of messages collected by a given amount of sensors. The last column provides the relative area covered by that number of sensors. \label{tab:messages}}
\end{table}

To demonstrate this fact with real flight data, we analysed more than 60 million ADS-B messages from aircraft at cruising altitudes (ca. 38,000\,ft) using 8 OpenSky sensors from an unplanned and crowdsourced deployment in Switzerland. Table \ref{tab:messages} illustrates the number of messages that are picked up by a given number of receivers. Only 14\% of all received messages are seen by 3 or more sensors on the ground and can be used for MLAT. If we take into account GDOP, we are left with only 5.24\% of usable messages. 

If we look at the area covered, we find that in our deployment the MLAT coverage makes up roughly 17\% of the overall covered area. The area where MLAT is reliably accurate is even smaller, only 6.67\% of the total area is covered by messages with a GDOP of less than 10. Thus, the usability of the received ADS-B messages for MLAT is reduced by a factor of more than 2.5 compared to localization methods that do not suffer from GDOP.



\section{Designing a New Aircraft Localization Approach for Crowdsourced Deployments}\label{sec:design}

We propose a grid-based k-NN approach which does not suffer from dilution of precision and works accurately regardless of the system geometry. As laid out above, this characteristic increases the signal utilisation and overall coverage of a deployment, thus vastly reducing costs and improving the detection range and speed of new aircraft at the same time.

\begin{figure}
\begin{centering}
\includegraphics[width=0.7\columnwidth]{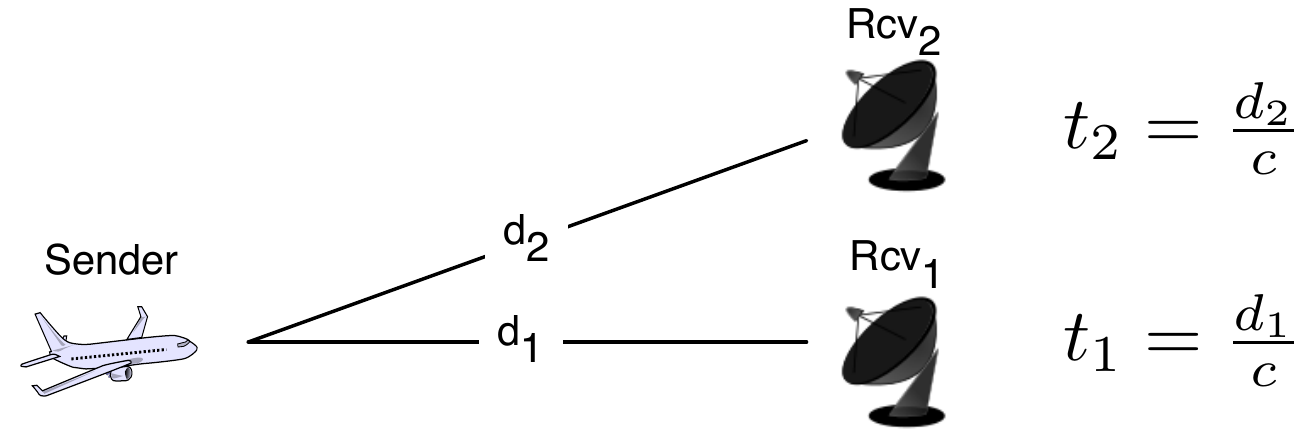}
\par\end{centering}

\protect\caption[An example illustrating the calculation of Expected TDOAs.]{An example illustrating the calculation of Expected TDOAs. The assumed distance of the sender to both receivers is divided by $c$. Subtracting the smallest time $t_{i}$ from the other times gives the TDoAs relative to receiver $i$. \label{fig:A-simple-example}}
\end{figure}

\subsection{Expected TDoAs}

The key insight of our approach is the use of \textit{Expected TDoAs}. As outlined before, the time differences of arrival of a received signal between multiple sensors are a physical primitive that can be used to establish the possible location(s) of a sender. For our approach, we extend this concept and (pre-)calculate the time differences that we would \textit{expect} to see based on any given location.
 
Concretely, we can use an idealized outdoor LoS propagation model suitable for the aircraft localization problem. We then calculate the absolute propagation times of an ATC signal's origin to the receiving ground stations by dividing the distances $d_{1},...,d_{n}$ between the sender and each of the stations $i=1,...,n$ by the speed of light $c$ (see Fig.\,\ref{fig:A-simple-example}).\footnote{As the propagation is not happening in a vacuum, this is an approximation, however, the difference is insignificant \cite{gomes2012next}.}

\begin{equation}\label{eq:toa}
t_{i} = \frac{d_{i}}{c}
\end{equation}

Subtracting the smallest resulting time $t_{min}$ from the other times provides the Expected TDoAs relative to the nearest receiver $i$:

\begin{equation}\label{eq:etdoa}
ETDoA_{i} = t_{i} - t_{min}
\end{equation}

We use this calculation of Expected TDoAs to estimate the the origin of the signal, as explained in the following section.

\subsection{Aircraft Localization with Expected TDoAs}

Putting these findings together, we design a novel approach to locate aircraft. In an offline training phase, a 2D grid of the surveillance area is computed that contains the expected TDoA measurements for each position for the given sensor deployment. In the online phase, for every incoming message, the k nearest neighbors of the messages' TDoAs on the grid are looked up. These neighbors are then averaged, resulting in the final estimate of the sender's location.\\

\begin{algorithm}[tb] 
\begin{algorithmic}[1] 
\STATE Input: $gridcoords$, $sensors$, $squaresize$
\STATE
\STATE $trainingset\gets$ [ ]
\STATE $grid\gets$ constructGrid($gridcoords$, $squaresize$)
\FOR{$\forall sensorcombinations$}
\STATE $TDoA\_training\gets$ [ ]
\FOR{$\forall  square$ $\in$ $grid$}
\STATE $TDoAs\gets$ computeTDoAs($sensors$.coords,$square$)
\STATE $TDoAtraining$.add($TDoAs$, $gridsquare$)
\ENDFOR
\STATE $trainingset$.add($TDoAtraining$, $sensorcombination$)
\ENDFOR
\caption{Localization offline phase. Requires coordinates of  sensors and grid as input and outputs the training sets for the online phase.}\label{alg:locestoff}
\end{algorithmic} 
\end{algorithm}

\begin{algorithm}[tb] 
\begin{algorithmic}[1] 
\STATE Input: $k$, $trainingset$, $flight$
\STATE
\LOOP   
\STATE $m\gets$ newPositionMessage($flight$)   
\STATE $r\gets$ receivers($m$)    
\IF{numberOfReceivers($m$) > 2}        
\STATE $TDoAs\gets$ calculateTDoAs($m$)
\STATE $trainingset\gets$ getTrainingset($r$) 
\STATE $knn\gets$ runKNN($trainingset$,$TDoAs$,$k$)
\STATE $estimate\gets$ getCenter($knn$)
\ENDIF 
\ENDLOOP 
\caption{Localization online phase. Requires the number of  neighbors $k$ and the $trainingsets$ from the offline phase as input and calculates the location estimate  of the message's origin. }\label{alg:loceston}
\end{algorithmic} 
\end{algorithm}

\paragraph{Offline phase}

Over an exemplary grid of $N\times M$ squares, we calculate the
fingerprint vector of expected TDoAs between the deployed sensors for every square using equations \ref{eq:toa} and \ref{eq:etdoa}. Based on this, we create the final training set by generating every subset of combinations with at least 2 sensors ($\sum_{i=2\,}^{n}{n \choose i},$ with $n$ being the number of sensors), e.g., 26 sets overall for a 5 sensor deployment. This is required when a new message is received by fewer than the maximum number of deployed sensors, with 2 sensors (or one TDoA measurement) as the lower limit. In that case the appropriate set is chosen during the online phase to find the $k$ nearest neighbors. Algorithm \ref{alg:locestoff} details the offline training phase.

\paragraph{Online phase}

In the online phase, new message data is analyzed and the location
determined (see Algorithm \ref{alg:loceston} for an overview of the whole process). Using the k-Nearest Neighbors algorithm, we obtain the closest points from our training grid that match the fingerprints of our test data. 

Setting the number of nearest neighbors to $k$, we match the received physical fingerprint $R=TDoA_{1},...,TDoA_{n}$ to the saved grid fingerprint $F$ based on their Euclidean distance

\begin{equation}
D_{(R,F)}=\sqrt{\sum_{i=1}^{n}(R\,_{TDoA_{i}}-\, F\,_{TDoA_{i}})^{2}}
\end{equation}

\begin{figure}
\begin{centering}
\includegraphics[width=0.95\columnwidth]{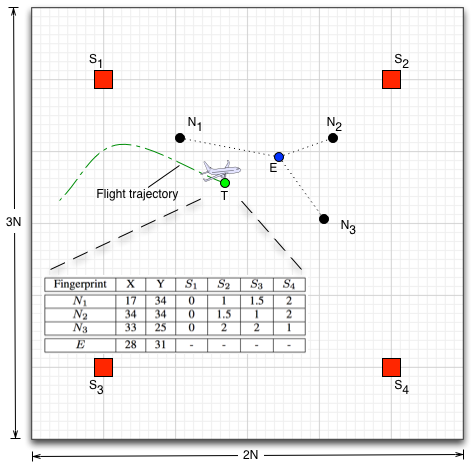}
\par\end{centering}

\protect\caption[Localization with 3-NN.]{Localization with 3-NN. Using TDoA data from 4 sensors $S_{1},...,S_{4}$, the 3 nearest neighbors $N_{1}$, $N_{2}$, $N_{3}$ in the lookup table are averaged to obtain the final estimate $E$. \label{fig:Location-estimation-using}}
\end{figure}

It is intuitive that in the spatial domain of our grid there are multiple neighbors that are approximately the same distance from our point of interest, hence $k$ is an important parameter influencing the accuracy. If $k>1$, the positions of all $k$ neighbors are averaged by taking the mean of the longitude and the latitude (see Fig.\,\ref{fig:Location-estimation-using} for an illustration). This constitutes the final estimate of the aircraft position, which on average is closer to the true location than any single neighbor as our evaluation will show.

\section{Experimental Setup}\label{sec:Experimental-Design}
 We will now describe our experimental setup and design, i.e., how we collected the data used to evaluate our approach, and the design of the grid used for the localization algorithm.

\subsection{Data Collection and Hardware}\label{sec:data-collection}

As ADS-B has been in the roll-out phase for years, we can use real-world data to estimate the propagation characteristics of ADS-B messages. We do not make any assumptions on hardware features such as sending power or antennas as there are many configurations found in different aircraft.

For our evaluation, we rely on real-world ADS-B data which we obtained from OpenSky a participatory sensor network for air traffic communication data, specifically ADS-B. It provides access to 3 years of historical raw message data as well as metadata, and offers a very fast query infrastructure, ideal for large-scale research projects. As of October 2016, it has saved more than 200 billion air traffic communication messages, covers about 3,000,000${km}^2$ on three continents, captures more than 16,000 unique aircraft every day, and has seen over 100,000 different aircraft overall. For more detail on OpenSky, its use cases for aviation security and privacy, and its big data infrastructure, please refer to \cite{Schaefer14,schaeferDASC2016,strohmeierDASC2015}.

For the present analysis, we use a dataset that spans the period between 26 June 2013 and 25 June 2014. This dataset contains more than 60 million ADS-B messages received from SBS-3 sensors manufactured by Kinetic Avionics. Besides the message content, they provide a timestamp of the message reception. From this data, we use 5 sensors that are closely located together to be able to calculate their TDoA data. The timestamps have a clock resolution of 50\,ns. The sensors have omnidirectional antennas and can receive signals from a distance of up to roughly 400\,km.

\subsection{Synchronization}

As our low-cost SBS-3 sensors do not provide built-in synchronization (e.g., via GPS), we synchronize our data a posteriori with the help of positional ADS-B messages sent by aircraft. By using the positional information in those messages and approximating their respective propagation time, we can recover the timing offset between our ground station sensors and achieve global synchronization. We also take into account the drift of the internal clocks to improve the results. Overall, this approach enables us to achieve synchronization that is low-cost and works well with minimal requirements. More accurate and efficient synchronization using GPS could help to further improve on the accuracy of our results. 

\subsection{Grid Design}

We construct a 2D grid over a typical flight altitude of 38,000\,ft (ca.~11,582\,m) with a size of 2 degrees longitude and 2 degrees latitude which, due to the Earth's spherical geometry, translates to an area of ca. $150\, km \times 220\, km\,=\,33,000\, km^{2}$. We obtain evenly-spaced approximate squares where the number of squares (or the squares' size) is a trade-off between performance and accuracy as elaborated in the evaluation section. Of course, computation time and accuracy also depend on the size of the surveillance area. 33,000\,km$^{2}$ are representative for wide area ATC surveillance, covering aircraft's en-route flight phase at cruising altitude.

\section{Evaluation}\label{sec:Evaluation}

\begin{table*}[t]
\noindent \begin{centering}
\begin{tabular}{|c|c|c|c|c|c|c|}
\hline 
Horizontal Error {[}$m${]} & MLAT & 600\,m$^{2}$ Grid & 300\,m$^{2}$ Grid & 150\,m$^{2}$ Grid & 75\,m$^{2}$ Grid & 50\,m$^{2}$ Grid\tabularnewline
\hline 
\hline 
Mean & 199.46 & 171.01 & 134.37 & 122.31 & 118.14 & 116.454\tabularnewline
\hline 
Median & 91.87 & 140.38 & 98.60 & 84.92 & 80.38 & 78.63\tabularnewline
\hline 
RMSE & 334.47 & 225.51 & 198.14 & 190.29 & 187.31 & 185.79\tabularnewline
\hline 
99th percentile & 1306.70 & 902.08 & 870.18 & 870.61 & 841.33 & 835.63\tabularnewline
\hline 
\hline 
Relative comp. time & 62.3\% & 100\% & 399\% & 1599\% & 7272\% & 16375\%\tabularnewline
\hline 
\end{tabular}
\par\end{centering}

\protect\caption{Horizontal errors in different grid square sizes using k-NN vs. MLAT, with 5 sensors and $k=5$. k-NN shows a better mean accuracy than MLAT of up to 41\% in our dataset.\label{tab:Gridcomparison}}
\end{table*}

In this section, we use the collected flight data to verify our approach. We analyze the location accuracy by comparing it to the GPS data broadcast by aircraft using ADS-B and discuss the impact of the number of neighbors and the grid setup on its performance. We further compare it against an algebraic MLAT algorithm as described in \cite{andersen},\footnote{Algebraic approaches require more receivers, but provide better accuracy and lower computational resources than numerical models as discussed in  \cite{mantilla2015localization}.} to show its improved efficiency in the examined aircraft localization setting.

\subsection{Aircraft Localization Accuracy}

To ensure a baseline for the accuracy of the localization method, we compare it with the GPS-based ADS-B position claims of legitimate flight data. We use a part of the whole dataset (comprising over 100,000 positional ADS-B messages), where every message has been seen by 5 sensors, providing us with sufficient TDoA measurements for our comparative analysis. All location claims are within the pre-defined surveillance grid in terms of latitude and longitude, while their mean altitude is 11,148.8\,$m$ ($\sigma=687.59\, m$). 

Table \ref{tab:Gridcomparison} shows the localization quality using k-NN with squares of five different sizes over an area of 33,000\,$km^{2}$ with $k=5$. As expected, increasing the number of squares has a positive impact; the smaller the square, the more accurate location predictions become. For example, a reduction in grid square size from 600\,$m^{2}$ to 300\,$m^{2}$ improves mean accuracy by 37.5\%. This naturally comes with a trade-off as the computational time to run the k-NN algorithm increases linearly by 400\%. Overall, we found that 150\,$m^{2}$ provides a good trade-off between accuracy and performance.

\begin{figure}
\includegraphics[bb=50bp 0bp 720bp 281bp,clip,width=1\columnwidth]{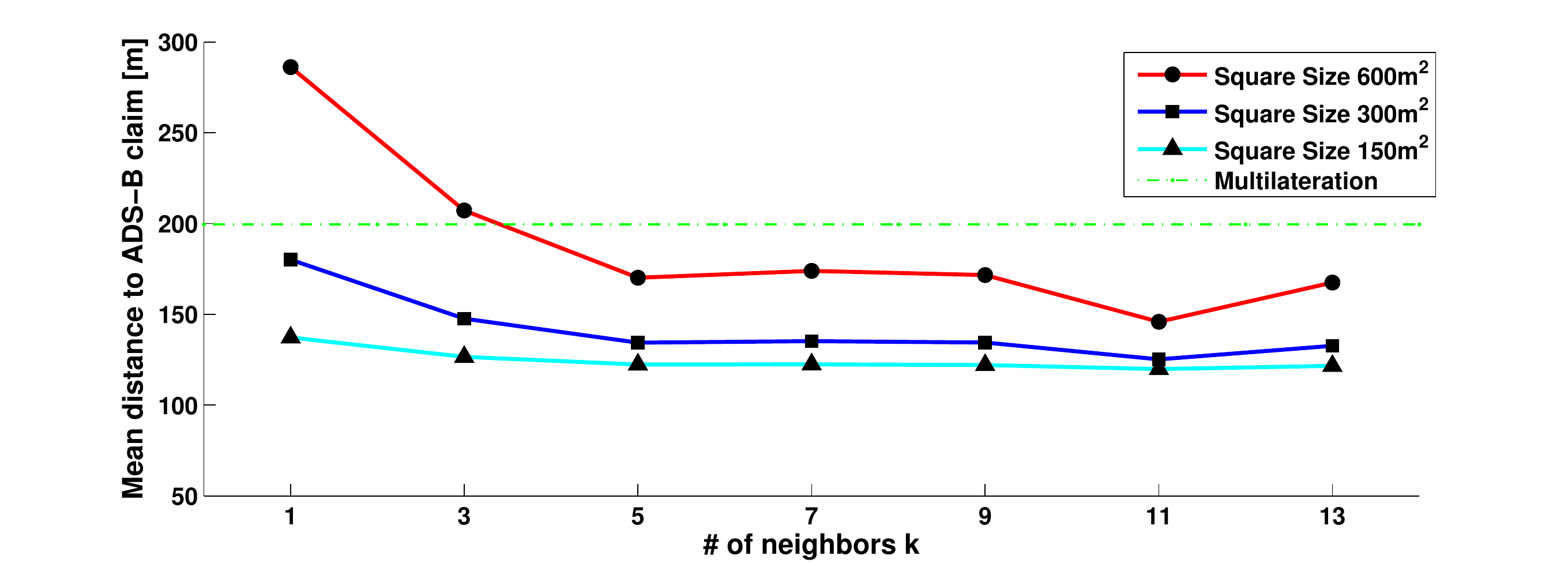}

\caption[Optimal choice of neighbors $k$ for different square sizes.]{Optimal choice of neighbors $k$ for different square sizes (MLAT as comparison). \label{fig:Optimal-choice-of}}
\end{figure}

Concerning the optimal choice of $k$, Fig.\,\ref{fig:Optimal-choice-of} illustrates the gains in accuracy when averaging a higher number of neighbors. We can see a large improvement until $k=5$ especially for a grid square size of 600\,$m^{2}$. Further decreases in mean accuracy are small and much less pronounced with smaller square sizes.

We compare k-NN with an algebraic MLAT algorithm using the same TDoA measurements from 5 sensors. The results show that with a 600\,$m^{2}$ grid size, k-NN does 14.2\% better than MLAT on mean errors, increasing to 41\% for a 50\,$m^{2}$ grid size. Overall, we find that k-NN does better than MLAT on noisy TDoA measurements such as those we experienced in our real-world data. Especially the more outlier-sensitive metrics RMSE and mean improve with k-NN while MLAT generally shows
good median results. Since k-NN does not suffer from dilution of precision, this is to be expected as the mean GDOP in our dataset is 24.35 ($\sigma=8.06$). Taking only ``good'' values below 10 into account, MLAT's metrics are bound to improve vastly. However, doing this also decreases the number of usable messages by over 90\%, reinforcing the fact that k-NN is useful in a much larger area.

\begin{table}
\noindent \begin{centering}
{\scriptsize{}}%
\begin{tabular}{|c|c|c|c|c|c|}
\hline 
{Error {[}$m${]}} & {\textbf{MLAT}} & {\textbf{2 Sens.}} & {\textbf{3 Sens.}} & {\textbf{4 Sens.}} & {\textbf{5 Sens.}}\tabularnewline
\hline 
\hline 
{Mean} & {199.5} & {26,956.7} & {311.8} & {147.3} & {122.3}\tabularnewline
\hline 
{Median} & {91.9} & {22,737.1} & {145.4} & {95.8} & {84.9}\tabularnewline
\hline 
{RMSE} & {334.5} & {33,380.4} & {761.3} & {237.6} & {190.3}\tabularnewline
\hline 
{99\%ile} & {1306.7} & {63,500.2} & {2,469.6} & {983.7} & {870.6}\tabularnewline
\hline 
\end{tabular}
\par\end{centering}{\scriptsize \par}

\protect\caption[Average horizontal errors using k-NN.]{Average horizontal errors using k-NN ($k=5$) with 150\,$m$ square size and different amounts of receivers. MLAT (5 sensors) is provided as comparison. \label{tab:Sensor comparison}}
\end{table}

The computational time is the trade-off for k-NN's accuracy and robustness. Only with the largest square size of 600\,$m^{2}$ it is comparable to MLAT. However, depending on the density of the airspace and the available equipment, even larger grids and longer computation times would not pose a problem in real-world settings.\footnote{The complexity of the algebraic MLAT algorithm is constant, while k-NN depends on the number of squares, i.e., both the size of the monitored area and the desired accuracy.} In scenarios where localization is used mainly to verify suspicious aircraft data, it is less critical as the examined amount of data is very small.

For our analysis, it is furthermore important to compare the impact of sensor numbers on localization. Table \ref{tab:Sensor comparison} shows the results for the same dataset and a 150\,$m^{2}$ grid size, if only a subset of the five sensors receives the messages. After analysing all possible subsets and averaging the results, we conclude that with only three sensors sufficient horizontal accuracy can be achieved.

\subsection{Robustness}

One of the main advantages of k-NN over MLAT is the fact that it is much more robust to noise. To analyze the impact of noisy signal measurements, we conducted simulations testing various noise levels against the two algorithms, using a scenario similar to the one with which we collected our real-world data using OpenSky. We randomly distribute 5 sensors on a 100x100\,$km$ grid, at altitudes randomly drawn from between 0 and 1,000\,$m$. We create 100,000 aircraft signals sent from the grid at a height between 10,000\,$m$ and 11,000\,$m$, calculate their ideal time of arrival at the 5 sensors and add white Gaussian noise $Z_{i}\thicksim\mathcal{N}(0,N)$ for each time $i$, where $N$ is the variance given in seconds. We repeat this simulation 1,000 times at each variance level $N$ to smooth out effects of the sensor placement.

\begin{figure}
\includegraphics[bb=90bp 0bp 1010bp 350bp,clip,width=1\columnwidth]{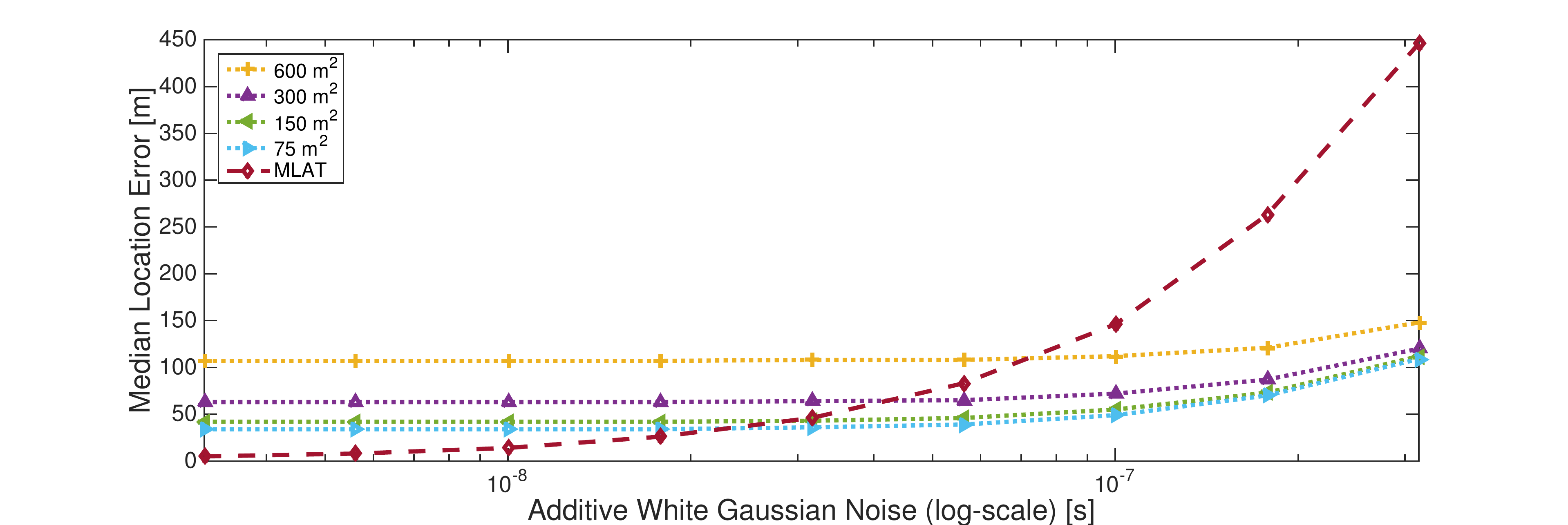}

\caption[Median location error depending on the level of noise $N$ affecting the measurements.]{Median location error depending on the level of noise $N$ affecting the measurements. k-NN results obtained with $k=5$ and 4 different square sizes.\label{fig:Median-location-error}}

\end{figure}

From the results shown in Fig.\,\ref{fig:Median-location-error},
we can conclude that the MLAT algorithm performs very well for no
or very little noise. This is not surprising for an exact method.
However, the algorithm is much less robust against increased noise
levels compared to k-NN. As the noise approaches a variance of $N=10^{-7.5}s$, we can see that the median location error of MLAT surpasses the one of k-NN with $k=5$ and 75\,$m^{2}$/ 150\,$m^{2}$ square sizes. This means that at a noise level of over $31.6ns$, our k-NN approach provides superior results. 

\begin{figure}
\includegraphics[bb=90bp 0bp 1010bp 350bp,clip,width=1\columnwidth]{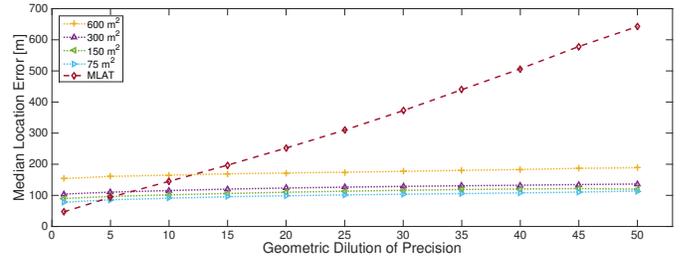}
\caption[Median location error depending on the geometric dilution of precision affecting the measurements.]{Median location error depending on the geometric dilution of precision affecting the measurements at a noise level of $N=10^{-7}s$. k-NN results obtained with $k=5$ and 4 different square sizes.\label{fig:Median-location-error-1}}
\end{figure}

Fig.\,\ref{fig:Median-location-error-1} gives further insight into the underlying reasons for the performance difference between both algorithms. It shows the median location error depending on the GDOP
of the measured signal. At a noise level of $N=10^{-7}s$, we find that MLAT only performs well when GDOP is extremely low, while k-NN is unaffected by this problem. Comparing MLAT to k-NN with various grid densities, we can see that only signals with GDOP < 5 can provide a low average localization error; higher GDOP levels are quickly outperformed by all square sizes. Unfortunately, such a low dilution of precision is present in only a very small fraction of the potential surveillance area (as illustrated in Fig.\,\ref{fig:dop-map}).

These findings illustrate that the MLAT algorithm requires extremely tight and costly synchronization and is still severely inhibited by the geometry of its receiver's locations. In contrast, k-NN with expected TDoAs can provide effective localization with good quality even with low-cost hardware such as Kinetic Avionics SBS-3 boxes and fully arbitrary receiver placement. These characteristics are very helpful not only in cases where deployment options are limited. They can further facilitate the use of TDoA-based localization in modern crowd-sourced receiver networks or pre-existing ADS-B deployments, both of which are regularly not optimized for perfect MLAT system geometry. It has to be noted that in professional ATC environments, continuous tracking algorithms further improve on the quality of the discrete localization results \cite{blackman1986multiple}, however, this can be applied to all localization techniques equally.


\section{Data Verification}\label{sec:attack-detection}

We now consider the verification of ADS-B data using our approach. Data verification can be necessary for several reasons, such as transponder malfunctions, inaccurate aircraft instrument data, or the deliberate broadcast of false data. Without loss of generality, we assume such an adversarial case, where false location data is injected into the ATC system by using manipulated ADS-B messages. 
The attackers have two different mobility models, influencing the credibility of their positional claims. The data  verification itself is done via a threshold: if an aircraft's ADS-B claim is deviating too much and for too long from the estimate, it is flagged as an anomaly, and thus as a possible attack.

\subsection{Test Data}

We use our real-world flight data obtained from OpenSky to test our data verification scheme. Out of the whole dataset described in \ref{sec:data-collection}, we analyze 1,341 legitimate flights and show that they are accurately verified by our system. Furthermore, we use data from two simulated attacker types (due to ethical and legal reasons, we do not implement real-world attacks) on the ground and in the air and check whether they will be verified or not. Each attacker injects 200 messages with the legitimate coordinates of a real flight from our sample and follows specific location patterns:\\

\begin{table}
\begin{centering}
\begin{tabular}{|c|c|}
\hline 
Attacker Type & Distance from claim {[}start/end/avg{]}\tabularnewline
\hline 
\hline 
Ground, mobile & 74.897\,/\,88.682\,/\,77.535\,km\tabularnewline
\hline 
Aircraft & 0\,/\,27.778\,/\,7.191\,km\tabularnewline
\hline 
\end{tabular}
\par\end{centering}

\protect\caption{Averaged horizontal distances from the two attackers' real positions to their claimed aircraft positions during the time that flight data is injected. \label{tab:Real-attacker-positions} }
\end{table}

\begin{itemize}
\item \textbf{Attacker 1} is ground-based and mobile, defined by a random horizontal start position on the grid, a random horizontal direction, and an altitude between 0 and 500\,m, moving on the ground with a speed of 50\,km/h.
\item \textbf{Attacker 2} is a legitimate aircraft, broadcasting a wrong track. Its starting position is the same as the real aircraft but then real and fake track diverge horizontally at a random angle between 10 and 45 degrees (at the aircraft's cruising altitude), making attacker 4 the most difficult to detect.\\
\end{itemize}

To illustrate their relative positions, Table \ref{tab:Real-attacker-positions} shows the average deviations from the real track for the two attacker types (at the beginning and the end of the attack, and on average, respectively). The attackers' TDoAs are calculated by dividing the 3D distance between the sensors by the speed of light $c$ and adding white Gaussian noise analogous to our real data to account for measurement and processing errors. We test each scenario 1000 times and analyze the detection rate.

\subsection{Evaluation}

We now evaluate the optimal verification thresholds based on the legitimate flights and derive the speed of detection against both attacker models. We further analyze how accurately the true origin of a signal can be located by our localization approach and MLAT.

\begin{figure}
\includegraphics[bb=75bp 0bp 880bp 360bp,clip,width=1\columnwidth]{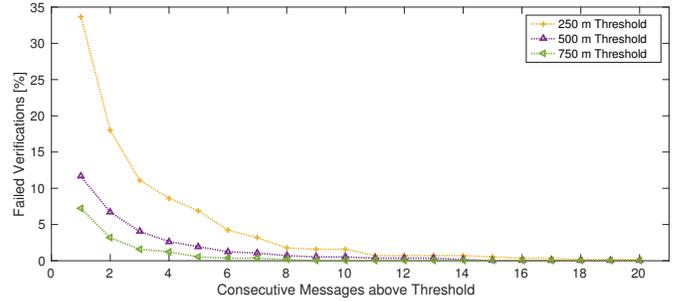}
\caption[Verification]{Analysis of different thresholds. The graph shows the percentage of real flights that fail to verify as legitimate based on distance from the aircraft location claim and number of consecutive messages exceeding this distance. k-NN results obtained with $k=5$ and $150\,m^2$ square size.\label{fig:verification}}
\end{figure}

Through experimental analysis of the legitimate OpenSky data, we first obtain a threshold that has not been exceeded by any of the legitimate flights in our dataset. Fig.\,\ref{fig:verification} shows three different thresholds (250\,m, 500\,m and 750\,m distance between our localization and the aircraft's position claim) and their associated likelihoods of verifying a real flight as legitimate. We find that our system should flag a given flight as illegitimate when the average deviation between the position claim and k-NN estimate exceeds 500\,m over a period of 15 messages or alternatively 9 messages with 750\,m. With these settings we encountered zero false positives in our test data, yet detect all false-data injections by attacker type 1 within these 9 (or 15) messages as their location always far exceeds the threshold. Attacker 2, who starts out from the correct position of the impersonated aircraft, exceeds the 500\,m threshold on average after 25 message and is thus still detected in fewer than 40 messages (or 38 for the 750\,m setting). This means, detection is possible after about 20 seconds without any lost messages or 40 seconds assuming a typical 50\% message loss on the ADS-B channel \cite{Strohmeier14}. Naturally, the precise thresholds depend on equipment and scenario and should be trained and fitted accordingly.

Besides the simple verification of flight data, localization methods can provide an estimate of an attacker's current location. Table \ref{tab:attacker-detection} provides the results of this estimation for both attacker types. k-NN with expected TDoAs not only accurately detects the distances between the attacker and the claim but can also give a good guess about the real origin of the signal. The horizontal estimate for the origin of message signals fits within approximately 2,000\,m of the real location for the ground-based attacker type. For the aircraft attacker type, we obtain an estimate that is accurate within the typical range for legitimate flight localization as discussed in the previous section, i.e. with an error of less than 200\,m.

\begin{table}
\noindent \begin{centering}

\begin{tabular}{|c||c|c|}
\hline 
&  \multicolumn{2}{c|}{Distance to attacker {[}km{]}}\tabularnewline
\hline 
Attacker Type  & k-NN & MLAT\tabularnewline
\hline 
\hline 
Ground, mobile & 1.918 & 44.947\tabularnewline
\hline 
Aircraft & 0.145 & 0.270\tabularnewline
\hline 
\end{tabular}
\par\end{centering}

\protect\caption[Mean distances between estimates and claimed location injected by an attacker \& mean distances to actual horizontal location of an attacker.]{Mean distances to actual horizontal location of an attacker. k-NN ($k=5$) with 150\,m square size. \label{tab:attacker-detection}}
\end{table}

In comparison, Table \ref{tab:attacker-detection} further illustrates the performance of MLAT in the same scenario. While MLAT can also verify the data and detect deviations between the signal origin and the position claim, it can not provide an accurate guess of the ground-based attackers type. Caused by large GDOP values of 1000 and more, the sensors' estimates of other objects in the same plane (i.e., on or near the ground) suffer from an error of more than 44\,km on average, making them essentially unusable. The origin of the aircraft attacker 2 is estimated with a low error of 270\,m, which is again in line with MLAT's performance for the localization of legitimate flight data in our setup.

This shows that, while it is entirely feasible (though costly) to build a multilateration system with good accuracy even for large surveillance areas in high altitudes, it is difficult to provide the same level of accuracy on the ground with the same system, especially when using cheap and unplanned deployments.

\section{Conclusion}\label{sec:Conclusion}

We proposed a new method for the localization of aircraft based on the expected time differences of arrival and the k-Nearest Neighbor algorithm. We evaluated our scheme with real-world flight data from our large-scale sensor network OpenSky using only low-cost ADS-B sensors in an unplanned and crowdsourced deployment. We find that it outperforms the popular multilateration approach in terms of range and coverage. Since it does not suffer from dilution of precision, it is possible to increase the overall coverage by a factor of up to 2.5 in our examined setting.

In terms of accuracy, our results show that the mean aircraft location accuracy can be increased by up to 41\% in comparison with an algebraic MLAT algorithm. Furthermore, as it does not suffer from dilution of precision, our approach is also more robust and less susceptible to noisy environments and bad system geometry, making it a better choice for pre-existing and unplanned receiver deployments.

Finally, we compare our approach against MLAT for the verification of aircraft location claims. The increased coverage improves the verification radius and enables us to detect even sophisticated aircraft-based attackers within 40 seconds. Furthermore, contrary to MLAT, it is possible to detect the approximate location of ground-based attackers within a mean horizontal error of about 2,000\,m.

Further work will involve analyzing the capabilities of our approach when it comes to localization in 3D space. While a natural extension, 3D localization is significantly more difficult as the degrees of freedom are increased. Nevertheless, even a less accurate solution could be helpful for verification purposes, in particular compared to MLAT, which is generally not able to give any useful altitude approximations of aircraft \cite{Galati2005}.

\balance
{\scriptsize{}\bibliographystyle{IEEEtran}
\bibliography{references}
}
\end{document}